\documentclass[]{emulateapj}

\usepackage{rotating}
\usepackage{latexsym}
\bibpunct{(}{)}{;}{a}{}{,} 

\begin{document}

\journalinfo{Accepted for publication in The Astrophysical Journal.}

\title{Timing and spectral properties of the accreting millisecond pulsar SWIFT~J1756.9--2508}

\author{Manuel Linares\altaffilmark{1}, Rudy Wijnands\altaffilmark{1}, Michiel van der Klis\altaffilmark{1}, Hans Krimm\altaffilmark{2,3}, Craig B. Markwardt\altaffilmark{2,4}, Deepto Chakrabarty\altaffilmark{5}}

\date{}

\def\rem#1{{\bf (#1)}}
\def\hide#1{}

\altaffiltext{1}{Astronomical Institute ``Anton Pannekoek'', University of Amsterdam and Center for High-Energy Astrophysics, Kruislaan 403, NL-1098 SJ Amsterdam, Netherlands.}

\altaffiltext{2}{CRESST and NASA Goddard Space Flight Center, Greenbelt, MD 20771}
\altaffiltext{3}{Universities Space Research Association, 10211 Wincopin Circle, Suite 500, Columbia, MD 21044}
\altaffiltext{4}{Department of Astronomy, University of Maryland, College Park, MD 20742}
\altaffiltext{5}{Department of Physics and Kavli Institute for Astrophysics and Space Research Massachusetts Institute of Technology, Cambridge, MA 02139}


\begin{abstract}

SWIFT~J1756.9-2508 is one of the few accreting millisecond pulsars
(AMPs) discovered to date. We report here the results of our analysis
of its aperiodic X-ray variability, as measured with the Rossi X-ray
Timing Explorer during the 2007 outburst of the source. We detect
strong ($\sim$35\%) flat-topped broadband noise throughout the
outburst with low characteristic frequencies ($\sim$0.1~Hz). This
makes SWIFT~J1756.9-2508 similar to the rest of AMPs and to other low
luminosity accreting neutron stars when they are in their hard states,
and enables us to classify this AMP as an atoll source in the extreme
island state. We also find a hard tail in its energy spectrum
extending up to 100~keV, fully consistent with such source and state
classification.

\end{abstract}

\maketitle

\section{Introduction}
\label{sec:intro}

Nine years after the discovery of the first accreting millisecond
pulsar (AMP), the number and variety of this type of neutron star
low-mass X-ray binary (NS-LMXB) continues to grow. The idea that a
neutron star could be spun up by accretion torques in the bosom of a
LMXB \citep{Alpar82, BhatHeuv91} was confirmed in 1998, when the first
millisecond pulses were seen from a LMXB \citep{Wijnands98}. Besides
coherent pulsations, two types of quasi-periodic millisecond X-ray
variability have been seen in AMPs: twin kilohertz quasi-periodic
oscillations (kHz QPOs) in SAX~J1808.4--3658 and XTE~J1807--294
\citep[][]{Wijnands03,Linares05} and burst oscillations also in
SAX~J1808.4--3658 and in XTE~J1814--338
\citep[][]{Chakrabarty03,Strohmayer03}. Together, these phenomena
place important constraints on kHz QPO and type I X-ray burst models,
offering a new insight onto accretion physics \citep[][for a recent
review]{vanderKlis06}. As far as low frequency variability is
concerned, the fastest spinning AMP, IGR~J00291+5934, showed the
strongest broadband noise with the lowest characteristic frequencies
observed to date in a NS-LMXB \citep{Linares07}.

Even though AMPs have provided new input to better understand
accretion onto low magnetic field NSs, important questions about their
physics remain open. For instance, if the magnetic field of the
neutron star in AMPs disrupts and channels the accretion flow, thereby
producing the observed pulsations, it is not clear yet why the
strength or configuration of such magnetic field should differ
drastically from that of non-pulsating NS-LMXBs \citep[see][for a
possible explanation]{Cumming01}. In this context it is interesting to
note that in three systems millisecond X-ray pulsations have been seen
to appear in and disappear from the persistent emission, producing
predominant \citep[HETE~J1900.1-2455;][]{Galloway07} intermittent
\citep[SAX~J1748.9-2021;][]{Altamirano08} or very rare
\citep[Aql~X-1;][]{Casella08} episodes of pulsations. This implies that
the AMP within them is only active or visible during a relatively
small fraction of the time, which may provide a link with
non-pulsating NS-LMXBs.

On June $7^{th}$, 2007, a new X-ray transient was discovered
\citep{Krimm07a} with the burst alert telescope (BAT)
onboard {\it Swift}. Subsequent {\it RXTE} observations revealed that
this was the eighth discovered AMP, with a pulse frequency of
$\sim$182~Hz and an orbital period of $\sim$54~minutes
\citep{Markwardt07a,Markwardt07b, Krimm07c}. The outburst lasted about two
weeks and a possible infrared counterpart was identified
\citep{Burderi07}. No radio pulsations were found in quiescence at
8.7 and 1.4~GHz \citep[][]{Possenti07,Hessels07}.

In the present work we analyze the aperiodic variability of the source
and compare it with other AMPs and with atoll sources \citep[a
low-luminosity class of NS-LMXB, see][]{HK89}. Using both PCA and
HEXTE data we also measure its 2-200~keV energy spectrum. We are
thereby able to classify SWIFT~J1756.9-2508 as an atoll source and to
characterize its accretion state.
 
\begin{table}[h]
\center
\tiny
\caption{Log of the observations.}
\begin{minipage}{0.5\textwidth}
\begin{tabular}{l c c c c c c c}
\hline\hline
Set & ObsID & Date\footnote{Observation start date in  MJD} & Detectors\footnote{Average number of active detectors} &  Count Rate \footnote{Average and standard deviation of the $\sim$2-35 keV PCA count rate, including all active detectors and not corrected for background (which we estimate from set E to be $\sim$22.3 c/s/PCU; see Section~\ref{sec:data})} & 512-s PDS\footnote{Number of power density spectra extracted from the observation, each of them 512 s long} \\
 & & & & (c/s) & \\
\hline
A & 93065-01-01-02 & 54265.2 & 2.0 & 111.0$\pm$1.4 & 6  \\
 & 93065-01-02-00 & 54266.0 & 2.0 & 102.3$\pm$3.8 & 16  \\
\hline
B & 93065-01-02-01 & 54267.0 & 2.0 & 95.3$\pm$0.9 & 4 \\
 & 92050-01-01-01 & 54267.1 & 3.0 & 136.7$\pm$1.0 & 6  \\
 & 92050-01-01-00 & 54267.4 & 1.7 & 75.3$\pm$21.1 & 24  \\
 & 92050-01-01-02 & 54268.1 & 2.5 & 109.3$\pm$20.9 & 12  \\
\hline
C & 92050-01-01-03 & 54269.0 & 2.0 & 71.3$\pm$1.3 & 6  \\
 & 92050-01-01-04 & 54269.2 & 2.0 & 68.8$\pm$4.2 & 5  \\
\hline
D & 92050-01-01-06 & 54270.2 & 2.0 & 46.0$\pm$2.2 & 10 \\
 & 92050-01-01-07 & 54270.6 & 2.4 & 62.4$\pm$12.6 & 20  \\
\hline
E & 92050-01-01-10 & 54271.6 & 2.0 & 49.4$\pm$0.9 & 3  \\
 & 92050-01-01-11 & 54271.8 & 2.0 & 50.0$\pm$2.2 & 4  \\
 & 92050-01-01-12 & 54271.9 & 3.0 & 67.0$\pm$1.6 & 5  \\
 & 92050-01-01-09 & 54271.9 & 1.0 & 20.9$\pm$0.3 & 5  \\
 & 92050-01-01-08 & 54272.4 & 1.3 & 28.6$\pm$11.1 & 19  \\
 & 92050-01-01-13 & 54273.0 & 2.0 & 47.4$\pm$1.2 & 5  \\
 & 92050-01-02-00 & 54273.5 & 1.0 & 21.8$\pm$0.6 & 11  \\
\hline\hline
\end{tabular}
\end{minipage}
\label{table:obs}
\end{table}

\section{Observations and Data Analysis}
\label{sec:data}

We used all pointed {\it RXTE} observations of SWIFT~J1756.9-2508
(Table~\ref{table:obs}). The observations started on June $14^{th}$,
2007, and were all performed with an $\sim$0.3$^{\circ}$ pointing
offset in order to avoid contamination from the nearby
($\sim$0.9$^{\circ}$ away) bright source GX~5-1. After applying
standard screening criteria\footnote{See {\it
http://heasarc.nasa.gov/docs/xte/abc/screening.html}. A conservative
filter on the time after SAA passage ($>$15 min) and the ELECTRON2
rate ($<$0.1~c/s) was applied before extracting the energy spectra.}
we extracted energy spectra from PCA (using only PCU2) and HEXTE
(using cluster B) data and power spectra from PCA data, using all
active PCUs.

After checking that no strong source falls within the field of view of
HEXTE-cluster B\footnote{{\it According to HEXTErock:\linebreak
http://heasarc.gsfc.nasa.gov/cgi-bin/Tools/HEXTErock/HEXTErock.pl}},
we extracted source and background spectra from standard mode
data. The background spectrum was obtained by averaging both rocking
positions. We applied dead-time correction and created response
matrices taking into account the above mentioned pointing offset.

A first estimate of the PCU2 background was given by the PCA
background model for faint sources. Given that the source lies in the
direction of the Galactic bulge and from the observed flux in the last
part of the outburst (or ``tail'', set E, see below) it is clear that
there is additional (non-instrumental) background flux, most likely
originating in the Galactic ridge \citep[note that the source was not
detected by][with the {\it Swift}-XRT on June $21^{st}$, within set
E]{Campana07}. The presence of an iron line at $\sim$6.5~keV whose
energy, width and flux remained constant as the overall flux decreased
also suggests that the Galactic ridge is the main source of
contaminating flux. In order to take this into account and under the
assumption that both instrumental and astronomical backgrounds remain
constant during the observations, we obtained the average spectrum of
the tail of the outburst and subtracted it from every PCU2
spectrum. This final empirical background estimation is independent of
PCA background modelling. The iron line was not present in the
resulting spectra. We created response matrices for the
$\sim$0.3$^{\circ}$ off-axis pointing.

\begin{figure}[b]
  \begin{center}

  \resizebox{0.6\columnwidth}{!}{\rotatebox{-90}{\includegraphics[]{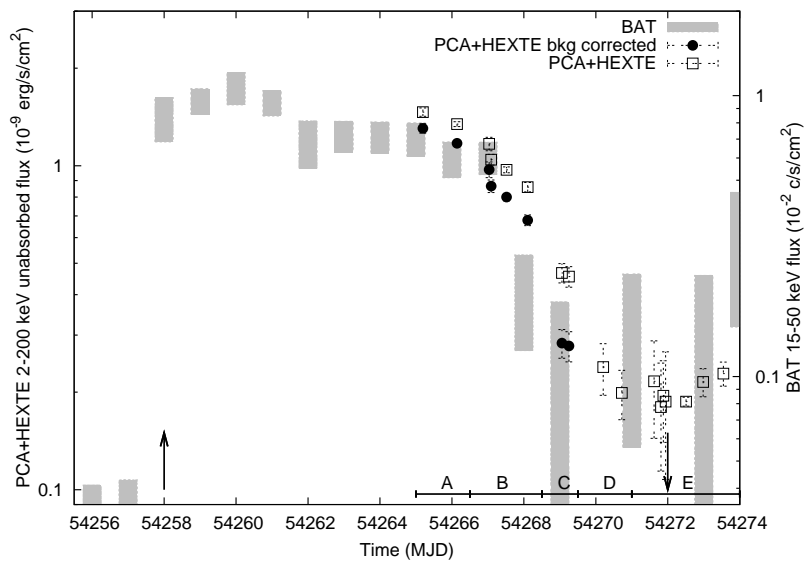}}}

  \caption{Light curve of the outburst of SWIFT~J1756.9--2508, showing 2-200~keV and 15-50~keV fluxes measured with {\it RXTE}-PCA\&HEXTE and {\it Swift}-BAT, respectively. Filled circles show fluxes after removing the Galactic ridge contribution (see Section~\ref{sec:data} for details), while for open squares only the instrumental background was subtracted. The upward arrow indicates the discovery date \citep[June $7^{th}$, 2007;][]{Krimm07a} and the downward arrow the date of the first non-detection with {\it Swift}-XRT \citep[June $21^{st}$, 2007;][]{Campana07}. Data sets A-E used in this work are shown with their corresponding labels along the bottom axis.}
    \label{fig:lc}
 \end{center}
\end{figure}

A detailed modelling of the energy spectra is beyond the scope of this
paper, and probably difficult due to the faintness of the source, so
we opted for the following phenomenological approach. We fitted the
joint PCA+HEXTE spectra with an absorbed blackbody plus power law
model using XSPEC v.11 \citep{Arnaud96}, allowing for a constant
factor (weighted mean 0.68$\pm$0.03) to account for the difference in
effective areas between PCA and HEXTE. The equivalent hydrogen column
density was fixed to 5.4$\times 10^{22}$ $cm^{-2}$, the value
measured by \citet{Krimm07c} from the {\it Swift} X-ray telescope
(XRT) spectrum. Figure \ref{fig:lc} displays the evolution of the
2-200~keV unabsorbed flux and the BAT 15-50~keV lightcurve, showing
the rise, decay, duration and peak flux of the outburst as well as the
data sets used in the timing analysis (see below).

In order to study the low-frequency variability, we performed fast
Fourier transforms (FFTs) on 512~s event mode data segments keeping
the original 125 $\mu$s time resolution. The power spectra thereby
cover the $\sim$0.02-4000~Hz frequency range. We used PCA channels 0
to 79 ($\sim$2.0-35.0 keV) excluding the highest
(background-dominated) channels to increase the signal-to-noise ratio,
and averaged all power spectra within each observation. No dead time
nor background correction was made before the FFTs and the Poisson
noise level was subtracted according to \citet{Kleinwolt04}
\citep[using the][formula and shifting the resulting level to match
the 2000--4000~Hz range, where no intrinsic power is
expected]{Zhang95}. When looking for changes in frequency and power of
the broadband noise (see Sec.\ref{sec:results}) we averaged
observations successive in time and with power spectra consistent
within errors in order to improve the statistics. The outburst was
divided into five contiguous data sets (labeled A-E; see
Table~\ref{table:obs} and Figure~\ref{fig:lc}). In set C low
statistics only allows us to constrain the break frequency and the
flat-top power level. In data set D the source was too faint to
measure accurately its energy or power spectrum (about 2.5 c/s/PCU),
and as discussed above data set E is background dominated. Power
spectra were ``rms-normalized'' \citep{vanderklis95b} using the count
rate observed in set E as background count rate.

We fitted the broadband power spectra with a sum of four Lorentzians
in ${\nu}_{max}$ representation
\citep{Belloni02}\footnote{$\nu_{max}=\sqrt{\nu_0^2+\Delta^2}$ gives
the characteristic frequency of the variability feature, where $\nu_0$
is the Lorentzian's centroid frequency and $\Delta$ its HWHM. The
quality factor $Q=\nu_0/2\Delta$ is a measure of the coherence of the
feature. Its strength is given by the integral power (0--$\infty$)
whose square root, in the normalization we use, is the fractional rms
amplitude of the variability.}, three of which were
zero-centered. Following previous notation
\citep{Belloni02,Straaten03,Linares07}, we call the different
Lorentzians $L_i$ and refer to their characteristic frequencies as
${\nu}_i$, where the label ``i'' identifies the variability feature
(see Section \ref{sec:results}).

\section{Results and Discussion}
\label{sec:results}

\begin{figure}[b]
  \begin{center}

  \resizebox{0.5\columnwidth}{!}{\rotatebox{0}{\includegraphics[]{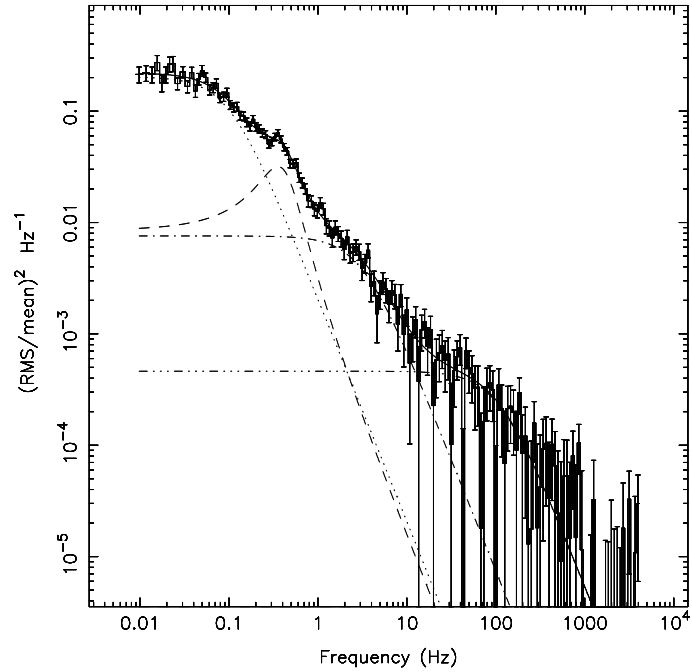}}} 

  \caption{Average power spectrum of sets A and B in Power vs. Frequency representation, together with the four Lorentzians function used in the fits.}
    \label{fig:ps}
 \end{center}
\end{figure}

Figure \ref{fig:ps} shows the broadband power spectrum of
SWIFT~J1756.9-2508. Its overall shape is typical of island and extreme
island states (IS/EIS) of atoll sources \citep[see
e.g.][]{Barret00,Belloni02}, as well as similar to low (luminosity)
and hard (spectrum) states of black hole X-ray binaries
(e.g. XTE~J1118+480; \citealt{Revni00}, XTE~J1550--564;
\citealt{Cui99}). Furthermore, the integrated (0.01-100~Hz) fractional
rms variability was very high ($\sim$35\%), also characteristic of
such low-hard (Comptonized) states. This allows us to classify this
AMP as an atoll source, something common to all the other AMPs known
to date \citep{Wijnands05}. The low characteristic frequencies of the
variability (see below) indicate that SWIFT~J1756.9-2508 was in the
EIS during its 2007 outburst \citep[see][for an overview of
states]{vanderKlis06}.

\begin{table}[h]
\center
\tiny
\caption{Parameters of the 4-Lorentzian fits to sets A and B.}
\begin{minipage}{0.5\textwidth}
\begin{tabular}{l c c c c r}
\hline\hline
 Parameter & $L_b$ & $L_h$ & $L_{\ell ow}$ & $L_u$ & $\chi^2$/d.o.f. \\
\hline
Set A & & & & & \\
\hline
${\nu}_{max}$ (Hz) & 0.12$\pm$0.01 & 0.45$\pm$0.03 & 3.1$\pm$0.9 & 53$\pm$30 & 167/150 \\
rms (\%) & 17.7$\pm$0.8 & 13.1$\pm$1.6 & 17.9$\pm$1.3 & 21$\pm$3 & \\
Q & 0 (fixed) & 0.9$\pm$0.3 & 0 (fixed) & 0 (fixed) & \\
\hline
Set B & & & & & \\
\hline
${\nu}_{max}$ (Hz) & 0.09$\pm$0.01 & 0.39$\pm$0.02 & 3.1$\pm$0.6 & 141$\pm$41 & 142/150 \\
rms (\%) & 18.3$\pm$0.8 & 15.1$\pm$1.5 & 21.0$\pm$0.9 & 34$\pm$3 & \\
Q & 0 (fixed) & 0.7$\pm$0.2 & 0 (fixed) & 0 (fixed) & \\
\hline\hline
\footnotetext{See Section~\ref{sec:data} for details of the fit function and notation. All errors are 1$\sigma$.}
\end{tabular}
\end{minipage}
\label{table:psfits}
\end{table}

\begin{figure}[b]
  \begin{center}

  \resizebox{0.6\columnwidth}{!}{\rotatebox{-90}{\includegraphics[]{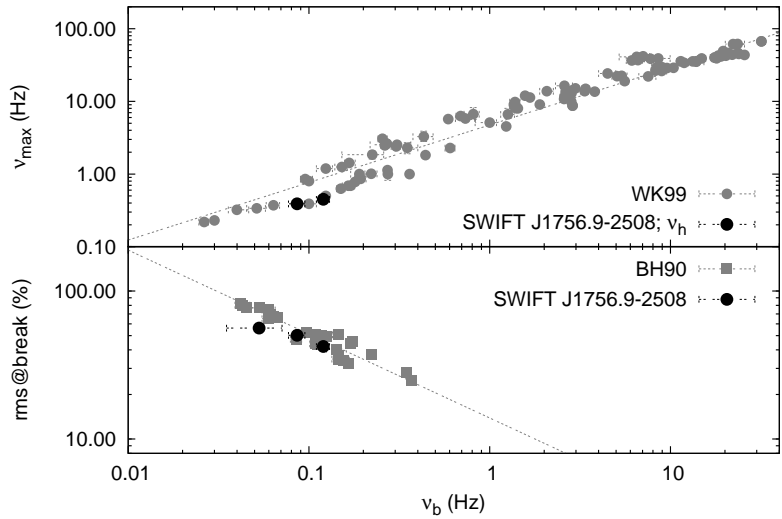}}}

  \caption{{\it Upper panel:} Correlation between the characteristic frequencies of $L_b$ and $L_h$, after \citet{WK99} \citep[see][for a discussion and extension of this correlation]{Belloni02}. {\it Lower panel:} Correlation between the frequency of the break in the power spectrum and the power level at the break frequency \citep{BH90}. SWIFT~J1756.9--2508 follows both of them (see Sec.~\ref{sec:results} for details).}
    \label{fig:bhwk}
 \end{center}
\end{figure}

The most prominent variability feature is a strong, band-limited and
flat-topped noise component ($L_b$, for ``break''), with a break or
cutoff at a frequency ${\nu}_b$$\sim$0.1~Hz. This break frequency lies
in the low frequency end of the range usually observed in NS-LMXBs. We
also detect a narrower (Q$\sim$0.8) component at higher frequencies
($\sim$0.4~Hz). The correlations between characteristic frequencies of
power spectral features are a useful tool to identify such features
\citep{WK99,PBK99,Straaten05}. The narrow component at $\sim$0.4~Hz
follows the correlation found by \citet{WK99} between $\nu_b$ and the
frequency of the ``hump'' ($L_h$) present above $L_b$ in both black
hole and neutron star systems (see Figure~\ref{fig:bhwk}), and we
therefore identify it as $L_h$. The next component in order of
increasing frequency has a characteristic frequency of 3.1~Hz, close
to (but a bit lower than) what we would expect from extrapolating the
relation between $L_{\ell ow}$ and $L_b$ found in atoll sources
\citep{Straaten02,Straaten03}. This leads us to tentatively identify
it as $L_{\ell ow}$. The highest frequency component that we find in
the power spectrum of SWIFT~J1756.9-2508 is most likely $L_u$, also
present also present in the same frequency range in IS/EIS of atoll
sources. A less likely candidate is the so-called ``hecto-Hertz''
Lorentzian, as that component seems to be present only when the
variability is at higher frequencies, with $\nu_b$ more than one order
of magnitude higher \citep{Straaten02}. All parameters from the fits
to sets A and B are shown in Table~\ref{table:psfits}.

In disk accreting X-ray binaries as the flat-topped noise shifts down
in frequency its flat top becomes higher. In other words, the rms
level at the break of the power spectrum is anticorrelated with the
frequency of that break. This was found by \citet{BH90} in Cyg~X-1 and
extended to other BH and NS systems by
\citet{Belloni02}. SWIFT~J1756.9-2508 also follows this
anticorrelation, as can be clearly seen from Figure~\ref{fig:bhwk}.

\begin{figure}[b]
  \begin{center}

  \resizebox{0.6\columnwidth}{!}{\rotatebox{-90}{\includegraphics[]{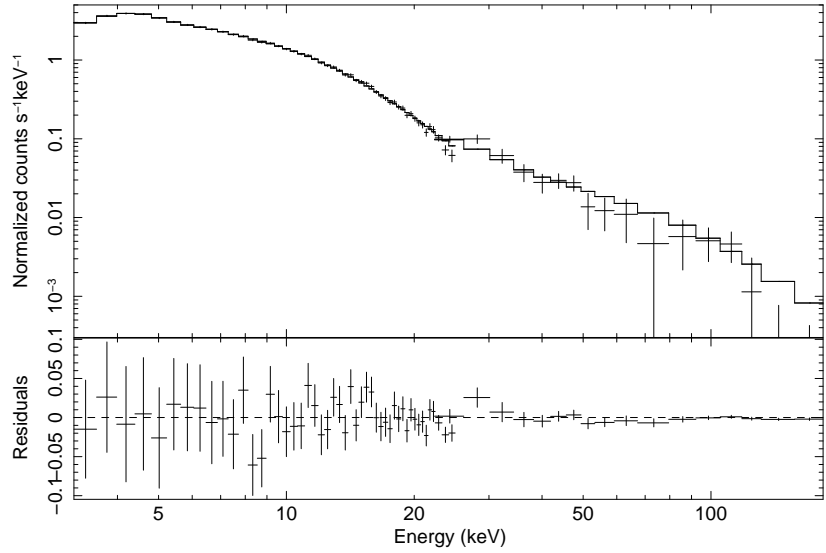}}}

  \caption{PCA+HEXTE energy spectrum of observation 93065-01-02-00, in set A, together with its fit function. Residuals are shown in the lower panel.}
    \label{fig:spec}
 \end{center}
\end{figure}

\begin{figure}[b]
  \begin{center}

  \resizebox{0.5\columnwidth}{!}{\rotatebox{0}{\includegraphics[]{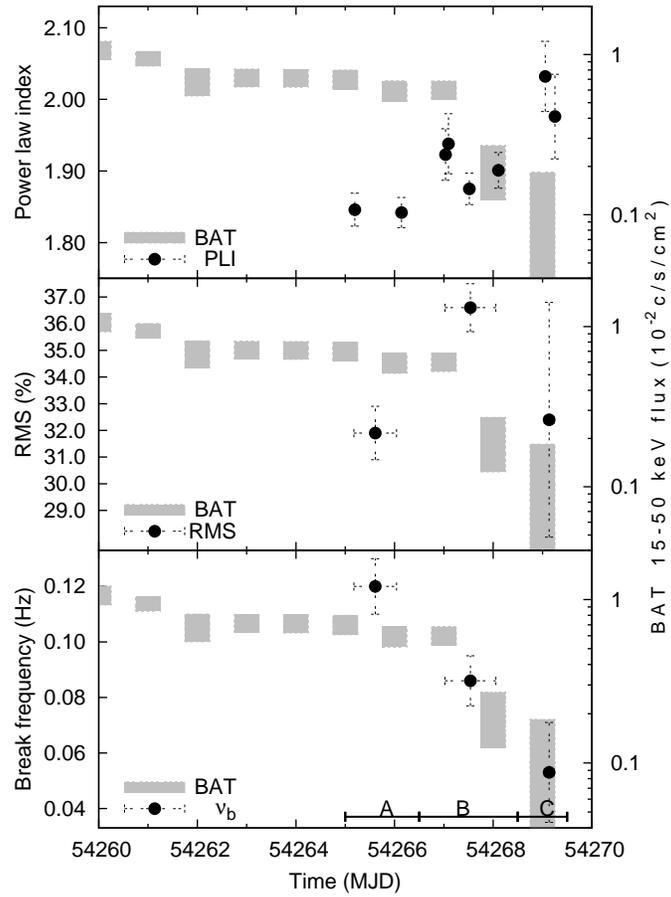}}}

  \caption{Time evolution of spectral index {\it (top)}, integrated (0.01-100~Hz) fractional rms {\it (middle)} and frequency of the break in the power spectrum {\it (bottom)}. The BAT light curve is shown in grey and the data sets used are indicated along the bottom axis.}
    \label{fig:tev}
 \end{center}
\end{figure}

The broadband 2-200~keV spectra from all observations within sets A, B
and C were reasonably well fitted (reduced $\chi^2 <$ 1.5 for 61
degrees of freedom) with the adopted (Sec.~\ref{sec:data}) model (see
Figure~\ref{fig:spec}). The resulting parameters are shown in
Table~\ref{table:spfits}. A hard tail extending up to $\sim$100~keV was
present in the energy spectra during most of the outburst, similar to
island states of other atoll sources \citep{DiSalvo04}. As can be seen
from Figure~\ref{fig:tev}, the resulting power law index increases
when the outburst decays. Several models have been proposed for the
emission process at work in the transition from outburst to
quiescence, including pulsar shock emission and thermal emission from
the neutron star \citep[][for a review]{Campana98}. Our data are not
sensitive enough to yield precise tests of such models, and in general
sensitivity limits and time constrains make this task a difficult one.

\begin{table}[h]
\tiny
\caption{Parameters of the spectral fits.}
\begin{minipage}{0.5\textwidth}
\begin{tabular}{l c c c r r}
\hline\hline
 ObsID & Ph.I. & P.L. & kT  & $\chi^2$/d.o.f. & 2-200~keV unabs.\\
  &  & norm &  (keV) &  & Flux ($erg/s/cm^2$)\\
\hline
Set A & & & & &\\
\hline
93065-01-01-02  &  1.85$\pm$0.02 &  (13.0$\pm$0.6)$\times 10^{-2}$ & 0.5$\pm$0.2  &  43/61 & (13.0$\pm$0.4)$\times 10^{-10}$ \\ 
93065-01-02-00  &  1.84$\pm$0.02 &  (11.0$\pm$0.6)$\times 10^{-2}$ & 0.7$\pm$0.1  &  53/61 & (11.7$\pm$0.2)$\times 10^{-10}$ \\ 
\hline
Set B & & & & &\\
\hline
93065-01-02-01  &  1.92$\pm$0.04 &  (11.0$\pm$0.9)$\times 10^{-2}$ & 0.5$\pm$0.1  &  59/61  & (9.7$\pm$0.5)$\times 10^{-10}$\\ 
92050-01-01-01  &  1.94$\pm$0.04 &  (11.0$\pm$1.1)$\times 10^{-2}$ & 0.6$\pm$0.4  &  61/61  & (8.7$\pm$0.4)$\times 10^{-10}$\\ 
92050-01-01-00  &  1.88$\pm$0.02 &  (9.6$\pm$0.5)$\times 10^{-2}$ & 0.6$\pm$0.1  &  70/61  & (8.0$\pm$0.2)$\times 10^{-10}$\\ 
92050-01-01-02  &  1.90$\pm$0.03 &  (9.1$\pm$0.6)$\times 10^{-2}$ & 0.7$\pm$0.1  &  83/61  & (6.8$\pm$0.3)$\times 10^{-10}$\\
\hline
Set C & & & & &\\
\hline
92050-01-01-03  &  2.03$\pm$0.05 &  (6.4$\pm$0.6)$\times 10^{-2}$ & 0.5$\pm$0.1  &  73/61  & (2.8$\pm$0.3)$\times 10^{-10}$\\ 
92050-01-01-04  &  1.98$\pm$0.06 &  (4.7$\pm$0.7)$\times 10^{-2}$ & 0.4$\pm$0.2  &  41/61  & (2.8$\pm$0.3)$\times 10^{-10}$\\ 
\hline\hline
\footnotetext{NOTE.-- In all fits the column density was fixed to 5.4$\times 10^{22}$ $cm^{-2}$ \citep{Krimm07c}. The power law (P.L.) is characterized by the photon index (Ph.I.) and the normalization, which corresponds to the photon flux density at 1~keV. In five cases the B.B. normalization was not significant but in all 8 observations it gave a $2-4~\sigma$ improvement in the fit, according to the f-test for additional terms. All errors are 1$\sigma$.}
\end{tabular}
\end{minipage}
\label{table:spfits}
\end{table}

We do not detect a significant change in the fractional rms of $L_b$
with energy in the studied energy range, suggesting that the spectrum
of this component is flat. Previous work has shown that in LMXBs the
X-ray luminosity does not uniquely determine the characteristic
frequencies of the variability \citep{Ford00,Straaten05,vanderKlis06}
so that another parameter must act to decouple luminosity and
variability frequencies. Figure~\ref{fig:tev} shows that as the
outburst decays and the inferred mass accretion rate drops the
fractional rms variability seems to increase, whereas the frequency of
the break decreases. The same behavior has been observed e.g. in
IGR~J00291+5934 \citep{Linares07}. A successful model for broadband
noise in LMXBs should explain both this short-term correlation between
luminosity and break (or ``cutoff'') frequency and the long-term
luminosity-frequency decoupling discussed above.

\textbf{Acknowledgments:}

{\it Swift}-BAT transient monitor results provided by the {\it Swift}-BAT team \newline ({\it http://swift.gsfc.nasa.gov/docs/swift/results/transients/}).

\clearpage

\bibliographystyle{aa}
\bibliography{biblio}

\end{document}